\documentclass[twocolumn,prl,showpacs,noshowkeys,superscriptaddress]{revtex4-1}

\usepackage{epstopdf}
\usepackage{amsfonts}
\usepackage{amssymb}
\usepackage{graphicx}
\usepackage{amsmath}

\begin{document}

\title{Gas Doping on the Topological Insulator Bi$_2$Se$_3$ Surface} 

\author{Mohammad Koleini}
\email{koleini.m@gmail.com}
\affiliation{Bremen Center for Computational Materials Science, University of Bremen, 28359 Bremen, Germany}
\affiliation{Hybrid Materials Interfaces Group, Faculty of Production Engineering, University of Bremen, 28359 Bremen, Germany}
\author{Thomas Frauenheim}
\affiliation{Bremen Center for Computational Materials Science, University of Bremen, 28359 Bremen, Germany}
\author{Binghai Yan}
\affiliation{Bremen Center for Computational Materials Science, University of Bremen, 28359 Bremen, Germany}

\date{\today}
\pacs{71.20.-b,73.20.-r,71.20.Nr}

\begin{abstract}

Gas molecule doping on the topological insulator Bi$_2$Se$_3$ surface with existing Se
vacancies is investigated using first-principles calculations. Consistent with experiments,
NO$_2$ and O$_2$ are found to occupy the Se vacancy sites, remove vacancy-doped
electrons and restore the band structure of a perfect surface. In contrast, NO and H$_2$ do
not favor passivation of such vacancies. Interestingly we have revealed a NO$_2$ dissociation
process that can well explain the speculative introduced ``photon-doping'' effect reported by recent
experiments. Experimental strategies to validate this mechanism are presented.
The choice and the effect of different passivators are discussed. This step
paves the way for the usage of such materials in device applications utilizing
robust topological surface states.

\end{abstract}

\maketitle

Three-dimensional (3D) topological insulators (TI) have attracted extensive research interest
recently~\cite{qi2010,moore2010,hasan2010,qi2010RMP}. Their novel topological surface states (TSS) in the bulk
energy gap open great practical potential in spintronics and quantum computation by realizing the
Majorana fermions~\cite{fu2008}. However, most TI materials available today are poor insulators in the bulk due to heavy defect
doping~\cite[][and references therein]{hasan2010,qi2010RMP}, hindering the utilization of the TSS in practice. 
To remove the bulk carriers, surface doping with gas molecules was adopted as a powerful tool in 
experiments~\cite{hsieh2009,chen2010b,xia2009b,hsieh2011,wray2011}, though the underlying detailed atomistic mechanisms are still unknown.
Exploring chemical reactions and understanding the modified surface structures are crucial to comprehend the current 
experimental results and help finding new methods to achieve the bulk insulating state.

As of today, the most attractive TI material is Bi$_2$Se$_3$~\cite{zhang2009,xia2009,hsieh2009}, 
demonstrating simple Dirac-type TSS along with large bulk energy gap.
Presumably due to the Se vacancies, however, it is found to be a $n$-doped semiconductor. 
To achieve a real bulk insulator, NO$_2$~\cite{hsieh2009,xia2009b,wray2011} and O$_2$~\cite{chen2010b,hsieh2011} 
gas species were used to dope the surface and successively remove the donated bulk conduction electrons.
In particular, the NO$_2$ doped Bi$_2$Se$_3$ surface exhibits a mysterious behavior, in which the surface loses
electrons when exposed to photon flux in angle-resolved photoemission spectroscopy (ARPES)
experiments~\cite{hsieh2009,xia2009b}. This is very different from the
cases of graphene~\cite{zhou2008} and another known TI material Bi$_2$Te$_3$~\cite{chen2009,xia2009b}, 
where the surface gains electrons back through desorbing NO$_2$ by the photon exposure. 
Though this photon-assisted stimulation or the so-called ``photon-doping'' method was employed to manipulate the surface bands in 
experiments~\cite{xia2010}, the underlying mechanism remains to be defined. In this letter, we
have studied from first-principles the reaction of gas molecules on the Bi$_2$Se$_3$ surface containing Se
vacancies and conclude how these affect the TSS.

\begin{figure}[htbp]
\includegraphics[width=\columnwidth]{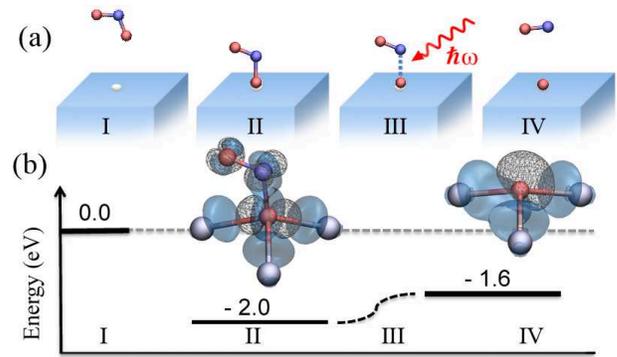}
\caption{(a) Schematic of NO$_2$ molecule adsorption and dissociation processes and (b) 
corresponding reaction energetics on the Bi$_2$Se$_3$ surface with a Se vacancy. 
The Se vacancy is indicated by the white dot (step I) and donates two electrons that can 
dope the surface. During NO$_2$ exposure, the molecule occupies the vacancy site quickly 
with one O binding to three Bi atoms from the second atomic layer (step II). 
The corresponding N-O bond become weaker compared to before adsorption (step I). 
The stable adsorption structure is shown in (b) together with charge density difference 
before and after adsorption. The donor electrons transfer partially from the vacancy to NO$_2$. 
Under external photon exposure (step III), the weakened N-O bond breaks, resulting in the NO$_2$ 
dissociation into NO + O (step IV). The NO molecule leaves the surface, while O passivates the vacancy. 
In this case, two donor electrons transfer to the O atom and the vacancy is compensated. 
White balls stand for Bi atoms, red for O, and blue for N. In the isovalue
surface plots of the charge difference, solid blue (wired gray) color denotes charge depletion (gain).}
\label{fig:fig1} 
\end{figure}

\begin{figure*}[htbp]
\includegraphics[width=2\columnwidth]{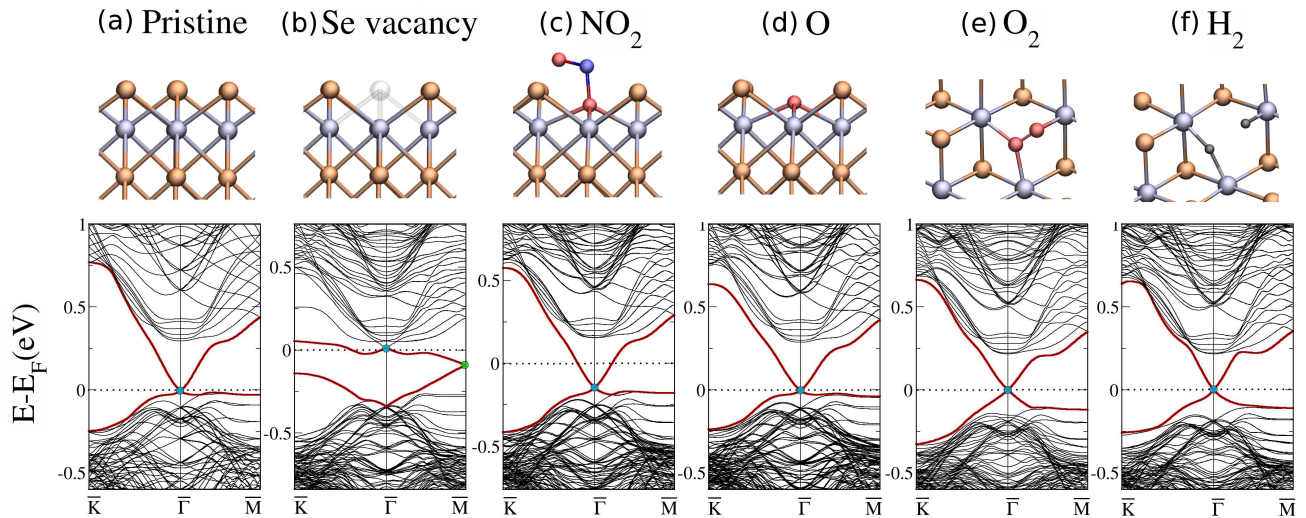}
\caption{ Atomic structures and band structures for (a) a perfect Bi$_2$Se$_3$ surface, (b) a surface with a
Se vacancy, (c) NO$_2$ adsorption, (d) O adsorption from the dissociated NO$_2$, (e) O$_2$
and (f) H$_2$ adsorption. Bands are aligned to their valence band maximum at $\overline{\Gamma}$.
The Fermi energy is shifted to zero.(a)-(d) show structures from
side view, whereas (e)-(f) from the top view. Red lines indicate the surface states. White
balls stand for Bi, yellow for Se, red for O, blue for N and small gray for H atoms. The
transparent ball indicate the missing Se at the vacancy in (b). Band structures are calculated from a 
$3\!\times\!3$ supercell of seven QLs thick. Colored circles denote Dirac points.} 
\label{fig:fig2} 
\end{figure*}

The first-principles molecular dynamic (FPMD) simulation was employed to investigate the dynamic
process of chemical adsorption and search for the stable atomic configurations. In bulk
Bi$_2$Se$_3$, five atomic layers form a quintuple layer (QL) while the coupling between two such QLs is
of the van der Waals type~\cite{zhang2009}. Here we used a two-QL-thick slab model containing
$3\!\times\!3$ primitive unit cells in the $xy$ plane to simulate a surface for the FPMD
adsorption studies. Thicker slabs were tested to yield correct and converged results. 
Both the pristine surface 
and the surface with a Se vacancy at the outmost layer were considered for
adsorption of NO$_2$, NO, O$_2$ and H$_2$ molecules. FPMD calculations including van der Waals
interactions~\cite{grimme2010consistent} have been performed at 300~K in microcanonical ensemble
(NVE) with an integration time-step of 1.0~fs~\footnote{In the system containing hydrogen the
time-step was decreased to 0.2~fs to cope with high frequency oscillations of H atoms.}. We let the
calculations run for 5.0~ps, while in worst case, roughly after 1.5~ps no more significant structural
changes were observed. Using the optimized structures from FPMD, the slab was extended to seven QLs thick to
prevent the interactions between states from the top and the bottom
surfaces~\cite{liu2010b,lu2010}. Inversion symmetry was maintained by placing the same molecules on
both surfaces. Subsequently, band structure calculations were carried out to investigate the electronic
properties.\footnote{FPMD calculations have been performed by \textsc{cp2k} code\cite{vandevondele2005} 
and electronic structure calculations have been performed by employing fully relativistic density-functional 
calculations including spin-orbit coupling (SOC) as implemented in the \textsc{openmx} code\cite{PhysRevB.72.045121}.
GGA-PBE for exch.-corr. \cite{Perdew1996} with grid cutoff of 250 Ry have been used.}

The results are shown in Fig.~\ref{fig:fig2}. The pristine surface has a single pair
of Dirac-like TSS with the Fermi level $E_F$ crossing the Dirac point at
$\overline{\Gamma}$, consistent with previous calculations~\cite{zhang2009}. When a Se vacancy
forms on the surface, $E_F$ shifts to the bulk conduction band bottom, resulting in
$n$-type doping as observed in experiments. Interestingly, the surface
states are energetically changed considerably by the existing vacancies. The original Dirac point
at $\overline{\Gamma}$ shifts upward and almost merges into the conduction band, while a new
Dirac point forms at the $\overline{M}$ point (Fig.~\ref{fig:fig2}(b)). In the bulk gap, $E_F$ still intersects
the TSS odd times, indicating the topological nontrivial character. It should be
noted that in our model Se vacancies distribute periodically on the surface, 
reflecting the employed supercell periodic boundary condition. However due to the vacancy
random distribution on the real sample surfaces, the ARPES measured only a single Dirac point
at $\overline{\Gamma}$ in the primitive Brillouin zone~\cite{xia2009,chen2009,hsieh2009}.
This illustrates a possible way to engineer the Dirac point by designing well-ordered surface potential.

\textit{NO$_2$ and NO.} In the FPMD simulations, NO$_2$ molecules are placed 
above both the pristine surface and the surface with vacancy. NO$_2$ moves freely above the pristine surface
whereby no strong binding effect is observed. The total energy calculation reveals
that NO$_2$ physisorbs on this surface with a weak adsorption potential energy (APE) of -~0.4~eV, 
resulting in tiny charge transfer from the surface to the molecule. 
However, on the defected surface NO$_2$ chemisorbed at the vacancy site quickly. 
One of the NO$_2$ oxygens, labeled as O1, adsorbs 
and binds to three Bi atoms from the second atomic layer, 
see Fig.~\ref{fig:fig1}(a) step~II. The equilibrium bond length of N-O1 is
elongated to 1.6~\AA\ compared to the other N-O bond of 1.3~\AA, indicating 
weakened N-O1 bond upon adsorption. Proceeding further, different FPMD replicas 
resulted into two situations.
Firstly, in the majority of the simulations, stable adsorption of NO$_2$ at the 
vacancy with the APE of -~2.0~eV took place. 
The corresponding band structure is depicted in Fig.~\ref{fig:fig2}(c), whereby 
the Dirac cone is recovered at $\overline{\Gamma}$. 
However, the surface is yet slightly electron-doped with	
$E_F$ 0.15 eV above the Dirac point. This indicates that NO$_2$ can not take away all donor electrons
from the vacancy. In the other situation, localized vibrations cause the weakened N-O1 bond to break.
Subsequently, NO$_2$ dissociates into NO which leaves the surface, while the O1 atom passivates the vacancy. 
The respective APE decreases to -1.6~eV. To estimate the
dissociation energy barrier, we performed density-functional calculations using the nudged
elastic band method~\cite{neb} and monitor dissociation to happen smoothly at a barrier
of about 0.4~eV. From the related band structure in Fig.~\ref{fig:fig2}(d), we see the O1 atom 
passivation shifts $E_F$ further down to the Dirac point and totally restores the band structure 
of the pristine surface. This can be understood by the fact that the single O1 atom can accommodate all two extra electrons from the Se vacancy, more than NO$_2$ in the former case, as illustrated by the
charge density change in Fig.~\ref{fig:fig1}(b). Contrary to NO$_2$, 
NO binds weakly (APE of -0.6~eV) and does not yield a similar good passivation 
of the vacancy, due to its much lower electron affinity.

\begin{figure}[htbp]
\includegraphics[width=\columnwidth]{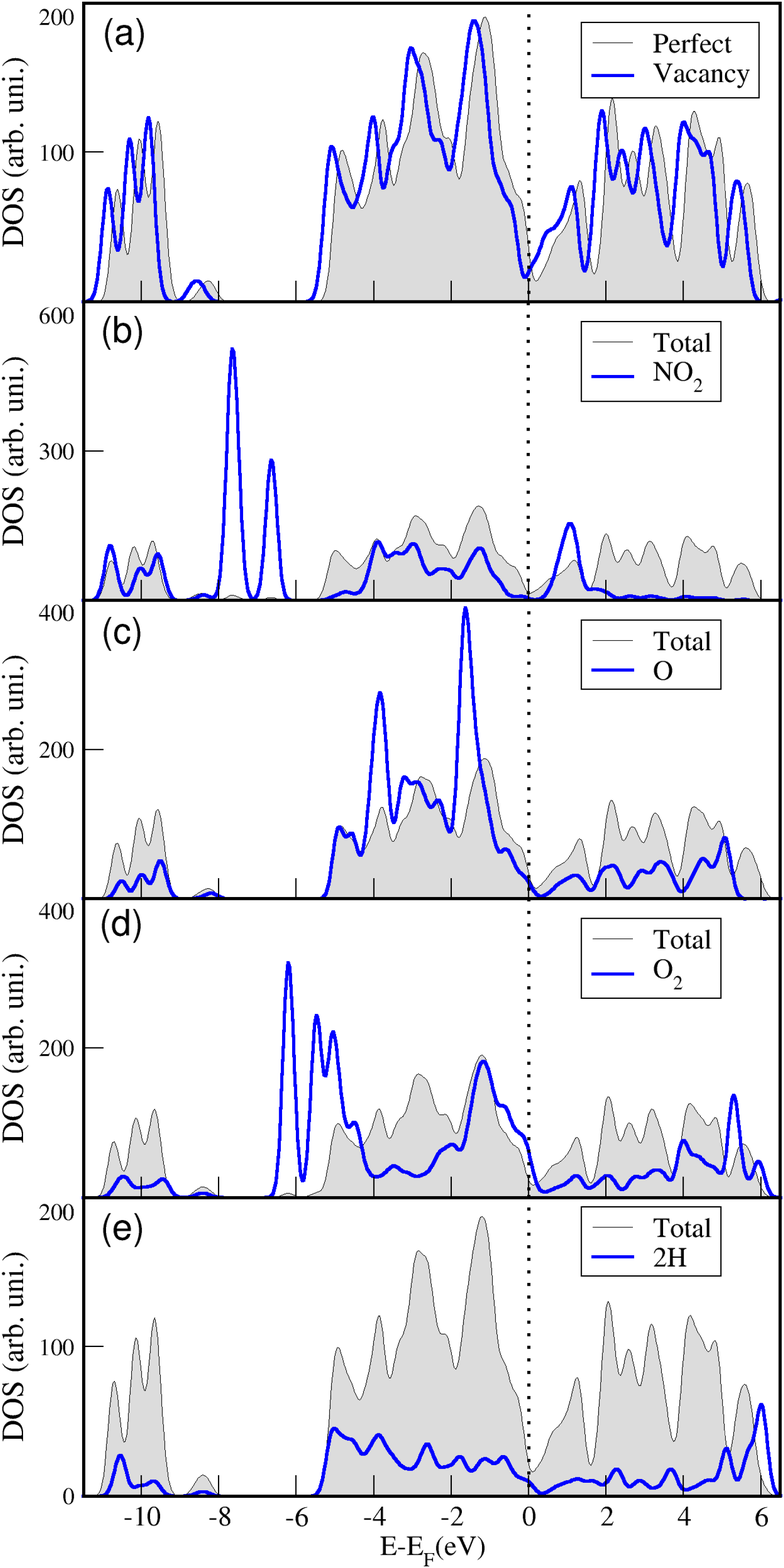}
\caption{Density of
states (DOS) for (a) the perfect pristine surface and the one with a Se vacancy, (b) NO$_2$, (c) a
single O atom, (d) O$_2$, and (e) H$_2$ passivated. To be more clear the adsorbates PDOS have been scales
by N$_\text{X}$/N$_{\text{Tot}}$, where N$_\text{X}$ is the number of atoms in the adsorbate and N$_\text{Tot}$
is the total number of atoms in the supercell. 
} 
\label{fig:fig3} 
\end{figure}
It is interesting to compare these results with recent ARPES
measurements~\cite{hsieh2009,xia2010}. In experiments, the Fermi level is 0.3~eV above the
Dirac point. Supposing all doped electrons are due to the Se vacancies, a dosage of
0.1 Langmuir(L) of NO$_2$ can shift $E_F$ down by 0.15~eV. After dosing with NO$_2$, in experiment 
two additional non-dispersive peaks appear at the binding energies of -4.0~eV and -7.5~eV~\cite{hsieh2009}. 
This agrees with our calculated density of states (DOS) projected
onto NO$_2$ (Fig.~\ref{fig:fig3}(b)), where partial DOS analysis (not shown here) indicates that 
the peak at $\sim$~-4.0~eV is mainly due to both O atoms and the other
at $\sim$~-7.5~eV due to all N and O atoms. The ARPES results are found to remain stable
with temperature varying from 10~K to 300~K. This can be explained by the large APE and the dissociation
barrier (0.4~eV), much too high to be overcome by RT thermal fluctuations. Thereby NO$_2$ will
neither desorb nor dissociate at 300~K. Moreover, the surface with a dose of 0.1~L
can also be further hole-doped by extensive photon (energy 28-55 eV) exposure until $E_F$
reaches the Dirac point~\cite{xia2010}, called as ``photon-doping''. Considering that NO$_2$ molecules 
dissociate into NO $+$ O under laser pulse irradiation~\cite{busch1972}, the above experimental
observation may be explained by the NO$_2$ dissociation process (Fig.~\ref{fig:fig1} step III). 
High-energy photons stimulate the N-O1 bond to break and subsequently transfer remaining electrons from the vacancy
to the O1 atom, finally downshifting $E_F$ to the Dirac point. We conclude that a dosage	
of 0.1~L of NO$_2$ passivates most of the Se vacancies on the surface, as the experimental
$E_F$ position (0.15~eV above the Dirac point) is the same as our calculated one, see
Fig.~\ref{fig:fig2}(c). Therefore photon exposure induces an O-passivated surface without extra donor
electrons. In addition, without photon exposure a dosage of 2.0~L more can also shifts
$E_F$ down to the Dirac point. This can be attributed to
the weak adsorption of NO$_2$ on the pristine surface, see above, where much less electron transfer happens compared 
to that of the strong adsorption on the vacancy.

The proposed dissociation mechanism can be easily checked in experiments. One way is to monitor
whether the composition of NO (NO$_2$) increases (decreases) after photon exposure. The
second way is to measure the valence band spectra. We predicted that the peak at $\sim$~-7.5~eV 
will become weaker or even disappear after extensive photon exposure, as indicated by
the projected DOS in Fig.~\ref{fig:fig3}(c). The third alternative is to \emph{see} the surface using STM or
from vibration spectra. 
Additionally, one can find that the monoatomic
oxygen, which is generally used in semiconductor industry for plasma ashing, can be an
excellent passivator to Se vacancies, if under controllable exposure.

\textit{O$_2$ and H$_2$.} The O$_2$ molecule also gets adsorbed in the vacancy site quickly with an
estimated APE of -~3.4~eV. The two O atoms still remain connectted to
each other with one O (labeled as O1) binding to one Bi atom and the other O (O2) binding
to the other two Bi atoms, as Fig.~\ref{fig:fig2}(e) shows. In the equilibrium structure, the O1-Bi bond length
is $\sim$~2.2~\AA, while two O2-Bi bonds form at $\sim$~2.4~\AA. Concluding from the
band structure analysis, O$_2$ captures two extra electrons from the vacancy and 
recover the neutral-charged TSS very well.
After adsorption no O$_2$ dissociation was observed in any of our MD simulations, clearly due
to stiffness of O=O bond.
This suggests why photon exposure had not any improving effect in experiments with O$_2$~\cite{chen2010b,hsieh2011}. 
In contrast, H$_2$ does not occupy the vacancy site in the MD
adsorption simulations. However after artificially saturating dangling bonds with hydrogens, the relaxed structure shows similar
bondings with three below Bi atoms as in the O$_2$ case, see Fig.~\ref{fig:fig2}(f). 
This passivation can also restore a perfect Dirac-type band structure
(Fig.~\ref{fig:fig2}(f)). Though H$_2$ adsorption is not favored in dynamics, single H atoms can be an ideal
passivation to Bi dangling bonds in theoretical studies, removing nontrivial surface
dangling bond states and simplify the physics underlying. 

In summary, we have studied the chemical adsorption of gas molecules (NO$_2$, NO, O$_2$ and
H$_2$) on the Bi$_2$Se$_3$ surface. The passivation of common Se
surface vacancies is found to gain the donor electrons much more effectively than weak adsorption on the pristine surface.
The local atomic structures after adsorption are revealed and the band
structures are compared. NO$_2$ is observed to passivate the vacancy and accommodate partially the vacancy-doped
electrons. By overcoming a moderate energy barrier ({\it e.g.} with laser stimulation), 
NO$_2$ dissociates into NO and a single O atom, whereby the latter occupies the vacancy and accommodates
all donor electrons. This dissociation and charge transfer can explain the mysterious
``photon-doping'' effect seen in ARPES experiments. As well, both O$_2$ molecule and the single O atom are found to passivate the vacancy 
very well and restore the surface band structure to the charge neutral state. 
In contrast, NO and H$_2$ are not favorable adsorbates on the Bi$_2$Se$_3$ surface.
The combinination of MD with electronic structure analysis is shown to be versatile in explaining 
experimental observations along with suggesting new routes for device engineering, 
by tailoring different dopants and TI materials.
\begin{acknowledgments}
B.Y. acknowledges the financial support from Alexander von Humboldt
Foundation in Germany. Computation time has been allocated at the University of Bremen.
\end{acknowledgments}

\end{document}